\pdfoutput=1

\documentclass[letterpaper, 10 pt, conference, hyphens]{ieeeconf}  

\IEEEoverridecommandlockouts                              
\usepackage[utf8]{inputenc}
\usepackage[T1]{fontenc}
\usepackage{amsmath}
\usepackage{amssymb}
\usepackage{bm}
\usepackage{epsfig}
\usepackage{graphics}
\usepackage{mathptmx}
\usepackage{hyperref}
\usepackage{mathtools}
\usepackage{xcolor}

\title{\LARGE \bf A Multi-Modal Respiratory Disease Exacerbation Prediction Technique Based on a Novel Spatio-Temporal Machine Learning Architecture and Local Environmental Sensor Networks
}

\author{Rohan Tan Bhowmik$^{1}$
\thanks{$^{1}$Rohan T. Bhowmik is a student at the Harker School, 500 Saratoga Ave, San Jose CA, USA 95138. He can be contacted at: {\tt\small 23rohanb@students.harker.org}}
}

\begin{document}

\maketitle
\thispagestyle{empty}
\pagestyle{empty}

\begin{abstract}

Chronic respiratory diseases, such as the chronic obstructive pulmonary disease and asthma, are a serious health crisis, affecting a large number of people globally and inflicting major costs on the economy. Current methods for assessing the progression of respiratory symptoms are either subjective and inaccurate, or complex and cumbersome, and do not incorporate environmental factors. Lacking predictive assessments and early intervention, unexpected exacerbations can lead to hospitalizations and high medical costs.

This work presents a multi-modal solution for predicting the exacerbation risks of respiratory diseases, such as COPD, based on a novel spatio-temporal machine learning architecture for real-time and accurate respiratory events detection, and tracking of local environmental and meteorological data and trends. The proposed new machine learning architecture blends key attributes of both convolutional and recurrent neural networks, allowing extraction of both spatial and temporal features encoded in respiratory sounds, thereby leading to accurate classification and tracking of symptoms. Combined with the data from environmental and meteorological sensors, and a predictive model based on retrospective medical studies, this solution can assess and provide early warnings of respiratory disease exacerbations.

This research will improve the quality of patients’ lives through early medical intervention, thereby reducing hospitalization rates and medical costs.
\end{abstract}

\section{\textbf{INTRODUCTION}}

Chronic respiratory diseases affect a large fraction of the world population, with Chronic Obstructive Pulmonary Disease (COPD) affecting 235 million and asthma affecting 339 million people worldwide, according to the World Health Organization [1]. Lacking effective early intervention, COPD and asthma cost over \$130 Billion annually in the U.S. alone [2].

Existing methods of diagnosis and tracking of these disease conditions in clinical practice, including widely-used patient questionnaires, are highly variable due to the subjectivity of definition, perception, and reporting of respiratory events. In fact, many respiratory diseases are often over- or under-diagnosed. Based on the study by Diab. et al, approximately 70 percent of COPD cases worldwide may be underdiagnosed, while 30 to 60 percent of those diagnosed with COPD may not have the disease at all [3]. As the treatment of respiratory diseases often requires the prescription of steroids, misdiagnosis can cause serious problems. 

Currently, no passive monitoring method exists for accurately predicting the exacerbation of respiratory conditions. A number of cough detection methods have been reported, but no accurate real-time tracking technique exists for passive and continuous monitoring. Commonly used methods involve subjective reporting, often leading to frequent and dangerous misdiagnosis [4-6]. Besides the respiratory conditions of the patient, environmental factors such as pollen, humidity, air quality, etc., also play a significant role in the disease progression, exacerbations, and hospitalizations [7]. However, currently there is no multi-modal predictive technique that incorporates the trends of both respiratory events and local environmental factors in order to assess the progression of the patient’s conditions.

Thus, the development of an accurate and real-time predictive solution for respiratory disease exacerbation that is easily accessible is highly needed, based on monitoring of patient’s respiratory events as well as the local environmental and meteorological parameters. The recent advances in connected devices, sensors, data technologies, and machine learning techniques present a significant opportunity to develop respiratory telehealth capabilities, allowing for accurate remote monitoring of patient conditions as well as assessing potential exacerbations with predictive Artificial Intelligence (AI) models.

This work presents a multi-modal solution for real-time COPD exacerbation prediction that includes a novel spatio-temporal artificial intelligence architecture for cough detection, real-time cough-count and frequency monitoring, analytics of the local environmental and meteorological factors utilizing data from sensor networks, and exacerbation prediction using both respiratory event tracking and environmental conditions based on retrospective medical studies. The goal of this research is to develop an early-warning system based on AI and multi-factor analysis to reduce hospitalizations and medical costs, and demonstrate the feasibility of deploying a passive, continuous, remote patient monitoring and telehealth solution for chronic respiratory diseases.

\section{\textbf{PRIOR RESEARCH}}

\begin{figure*}[thpb]
      \centering
      \label{table1}
      \includegraphics[width = 6in]{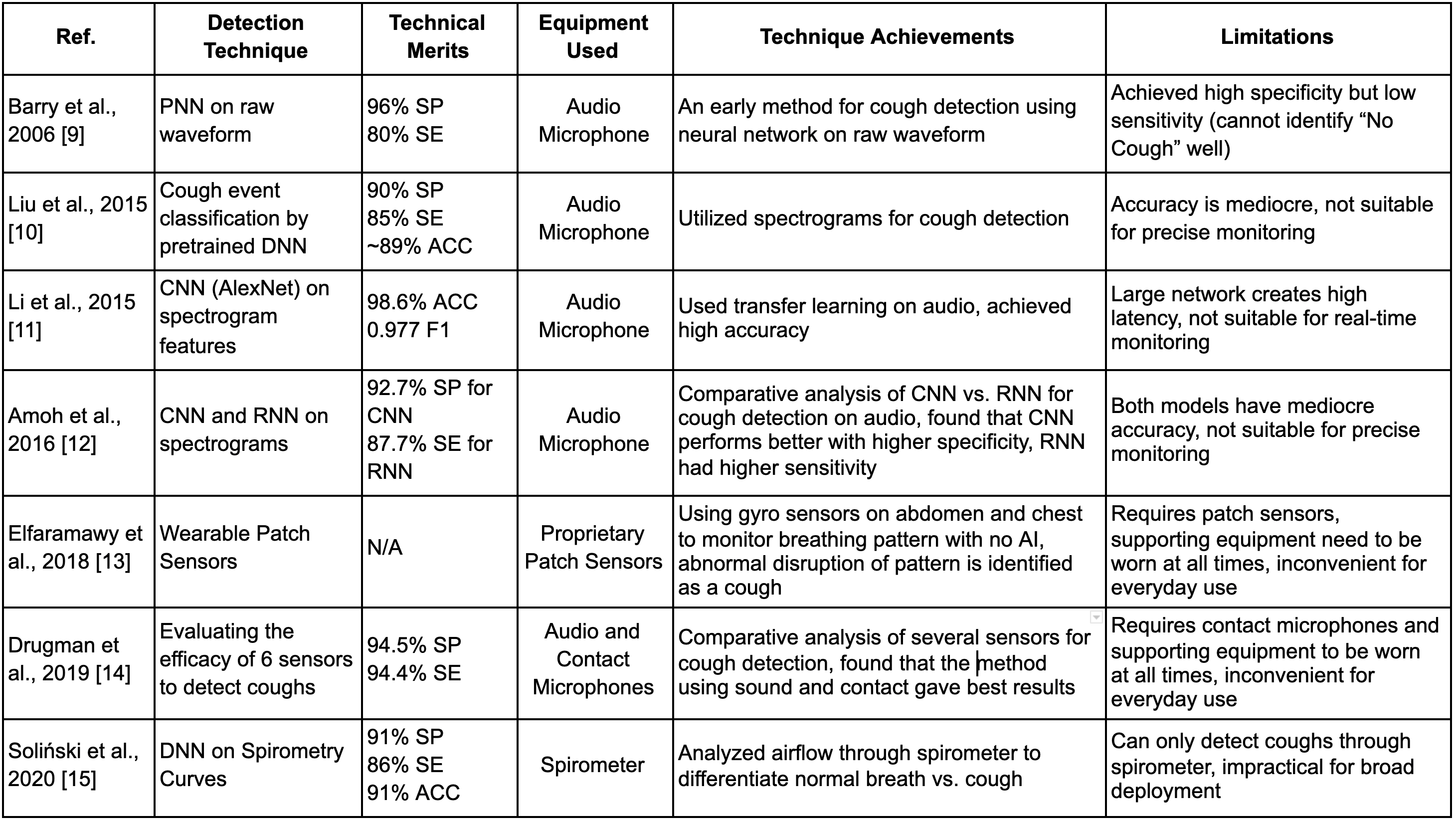}
      \caption{A survey of previously reported techniques for automatic cough detection (PNN = Probabilistic Neural Network; SP = Specificity; SE = Sensitivity; ACC = Accuracy). The above techniques follow these observations: i) generally higher accuracies were achieved with more complex models utilizing spectrograms; ii) techniques aided with extra equipment produced better results; iii) no single technique simultaneously meets all of the following requirements: highly accurate, efficient, passive and continuous monitoring, does not need extra equipment.}
\end{figure*}

Researchers have previously identified that monitoring a patient’s respiratory events can be utilized to assess the patient’s condition [8]. In order to automate this process, a number of cough detection solutions have been proposed [9-15]. A survey of previously reported techniques, performances and limitations are listed in \hyperref[table1]{Fig. 1}. Earlier methods used relatively simpler techniques, such as probabilistic statistical models on waveform data [9], but also yielded low accuracies. On the other hand, more recent studies have used specialized equipment and complex setups, such as wireless wearable patch sensors [13] or spirometers [15], to achieve relatively better results. However, no single technique simultaneously meets all of the following requirements: highly accurate, efficient, passive and continuous monitoring, and does not need extra equipment.

With the recent advancements in the field of artificial intelligence, researchers have moved towards exploring solutions based on Deep Neural Networks (DNN). Several researchers have demonstrated detection of cough with either Convolutional Neural Networks (CNN) or Recurrent Neural Networks (RNN). Traditional CNN models are based on learning and detecting spatial features in the data and are typically used for image-based analysis, whereas RNN models are based on extracting temporal features and are often used for time-sequenced tasks such as speech processing. Since respiratory sounds, when converted to spectrograms, encode key spatial and temporal signatures, neither of the traditional models is well suited for respiratory event classification.

Some researchers have recently reported combined Convolutional-Recurrent Neural Networks (CRNN) for acoustic analysis [16-20]. These CRNN models have been shown to work better than CNN and RNN in both image processing and sequence-related tasks [16, 19], but these frameworks do not fully utilize the spatial/temporal feature extraction capabilities of CNN/RNN architectures as they are created by simply stacking RNN layers after CNN layers in a sequential manner. The development of machine learning architecture based on deeply meshed spatio-temporal feature learning for respiratory sound classification has not been previously explored.

Medical researchers have also shown that several key environmental and meteorological factors are related to the exacerbations of COPD [7]; however, this research has not been combined with real-time monitoring of respiratory events to develop predictive models for exacerbations.

\section{\textbf{METHODS}}

\subsection{Proposed Multi-Modal System Architecture}

\begin{figure*}[thpb]
      \centering
      \label{fig1}
      \includegraphics[width = 6in]{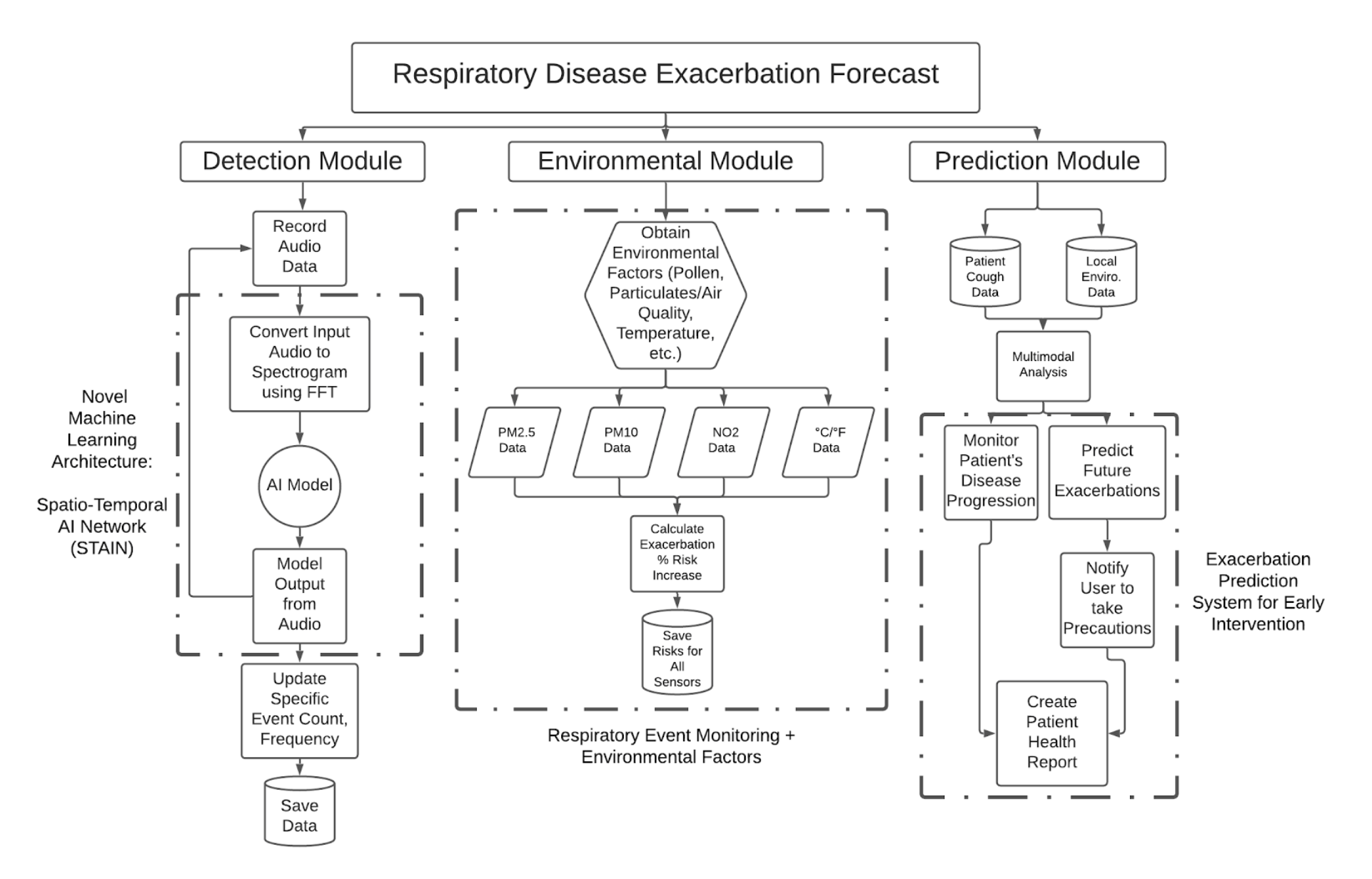}
      \caption{This flowchart represents the proposed system architecture for real-time multi-modal exacerbation prediction. The detection module depicts the respiratory event analysis system using a novel spatio-temporal artificial intelligence neural network. The prediction module depicts the disease exacerbation prediction system using the output of the machine learning model and environmental factors. The prediction module takes the respiratory event data and trends from the detection module, forecasts the progression of the patient's conditions, and provides necessary alerts for early intervention.}
\end{figure*}

In this project, a novel multi-modal COPD patient monitoring and exacerbation prediction system has been developed based on real-time analysis and tracking of both respiratory events and environmental factors. As shown in \hyperref[fig1]{Fig. 2}, the system architecture consists of three stages: i) a detection module, ii) an environmental module, and finally, iii) a prediction module. 

The detection module utilizes a new spatio-temporal machine learning algorithm for accurately detecting coughs from real-time audio and tracking the patient’s cough count and frequency. Simultaneously, the environmental module acquires local environmental and meteorological data from nearby weather stations and sensor networks to calculate the percentage increase of exacerbation risks in any location around the world based on the results of retrospective medical studies. Finally, the prediction module combines the historical cough count data and trends from the detection module and the calculated exacerbation risk increase from the environmental module in order to forecast the progression of the patient’s conditions, and alert the patients and caregivers for early interventions.

\subsection{Detection Module}

The detection module, as shown on the left-hand side of the system architecture diagram in \hyperref[fig1]{Fig. 2}, consists of a new AI model for real-time detection and tracking of cough. As described earlier, previously reported models for respiratory sound analysis are based on the traditional convolutional, recurrent, or the more recent convolutional-recurrent structures. In this project, a new machine learning algorithm has been developed that incorporates a novel hybrid framework by deeply meshing convolutional and recurrent architectures, enabling more efficient extraction and analysis of spatio-temporal features, leading to better accuracies for classifying and tracking respiratory events.

The following subsections describe the new spatio-temporal machine learning framework for classifying and tracking respiratory events, creation of the dataset to train and test the model, the results of benchmarking the proposed model with traditional neural network architectures, and a live demonstration application showcasing the capability of real-time classification of respiratory sounds.

\subsubsection{A New Machine Learning Architecture for Respiratory Sound Analysis}

\begin{figure*}[thpb]
      \centering
      \label{fig2}
      \includegraphics[width = 6in]{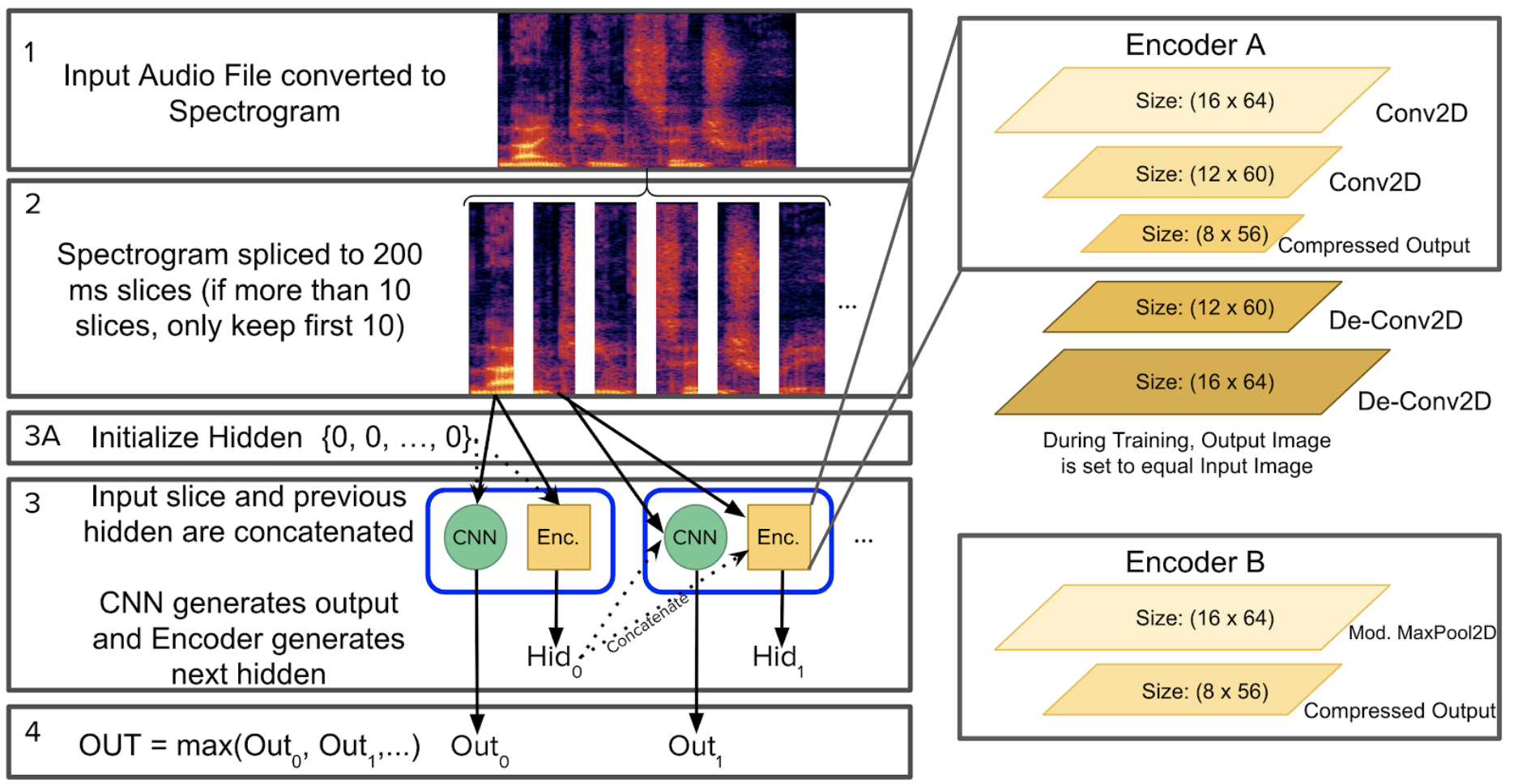}
      \caption{Architecture for the new machine learning model, which is referred to as the spatio-temporal artificial intelligence network (STAIN). This proposed AI model deeply blends the elements of both convolutional and recurrent neural networks, and effectively learns both spatial and temporal features encoded within the respiratory sound spectrograms for accurate classifications.}
\end{figure*}

The new AI model, henceforth referred to as the Spatio-Temporal Artificial Intelligence Network (STAIN), interweaves convolutional neural network models within a recurrent neural network architecture, allowing for sequential image analysis over the time domain. The architecture of the STAIN framework is shown in \hyperref[fig2]{Fig. 3}. First, the respiratory sound files are converted to corresponding spectrogram images by performing Fast Fourier Transforms. The resulting spectrogram is split into 200 millisecond slices, which are used as inputs for the machine learning model. 

As illustrated in \hyperref[fig2]{Fig. 3}, the machine learning model architecture incorporates a hybrid network based on a deep mesh integration of convolutional and recurrent architectures to learn spatio-temporal features. The STAIN framework consists of a CNN model which evaluates the corresponding audio slices and outputs its predicted confidence. The CNN architecture is a variation of Yann Lecun’s seminal LeNet model [21], which can flexibly adapt to any image dataset. Specifically, it consists of two groupings of Convolutional Layers of 2x2 kernels and 2x2 Maximum Pooling Layers followed by Rectified Linear Unit (ReLU) activation function. Then, the resulting data is flattened into a one-dimensional array before feeding it into two Fully Connected (Dense) Layers to reduce the number of neurons down to just one. The final output is then passed through a Sigmoid Layer to obtain a value between (0, 1).

The CNNs analyzing separate parts of the input image enable spatial feature extraction, while the Encoders passing down compressed inputs as RNN’s hidden variables enable temporal feature extraction. Various designs for the Encoder have been explored, starting with a simple architecture consisting of a single Maximum Pooling layer, shrinking the input into a hidden variable. A simple Variational Auto-Encoder (VAE) has also been created, consisting of two Deconvolutional Layers followed by Convolutional Layers.

Effectively, each slice of the spectrogram image is assigned to an RNN unit, wherein a CNN generates an output and the Encoder generates the hidden data. Each output represents the probability of a cough during that slice. The hidden outputs carry on information from previous slices and are concatenated to the next slice. The final output is the maximum of all the outputs from all slices. All the codes in this project were written in Python, and the machine learning models were implemented using the PyTorch Libraries.

\subsubsection{Creation of the Dataset}

In order to train and evaluate the proposed STAIN machine learning model as well as benchmark with other state-of-the-art models including CNN, RNN and CRNN, an augmented dataset of audio segments were created and partitioned into 10,000 training files with coughs, 10,000 training files without coughs, 1,000 testing files with coughs, and 1,000 testing files without coughs. The models were trained only on the 20,000 training files and tested only on the 2,000 testing files in order to objectively evaluate and compare the performance of various models.

First, roughly 500 cough sound files were downloaded from the Free Sound Database (FSD) from Kaggle’s audio tagging competition [22] and every file was adjusted to only contain either a cough burst or coughing fit. The cough files were sufficiently diverse, containing many variations of coughs from individuals of both genders and from a wide range of ages (from babies to elderly). Each file also has it’s unique recording quality, mimicking the varying degrees of audio quality from different devices.

In order to augment the data, the rest of the audio files from Kaggle’s FSD were utilized. To create an augmented audio file, an empty audio file is created with a duration randomly chosen between 2 seconds and 5 seconds. Then, using the PyDub Library, a randomly chosen number of non-cough files from the FSD are superimposed on the targeted augmented file. Each of the added audio files are placed at a randomly chosen timestamp, with audio exceeding the augmented files trimmed off. The result of this process creates an augmented audio file categorized as “No Cough”. To turn it into a “Cough” file, one of the cough files from the FSD is added in a similar fashion. Additionally, each added file’s decibel gain is randomized to simulate sounds from varying distances.

\subsubsection{Benchmarking and Results} 

\begin{figure*}[thpb]
      \centering
      \label{table2}
      \includegraphics[width = 6in]{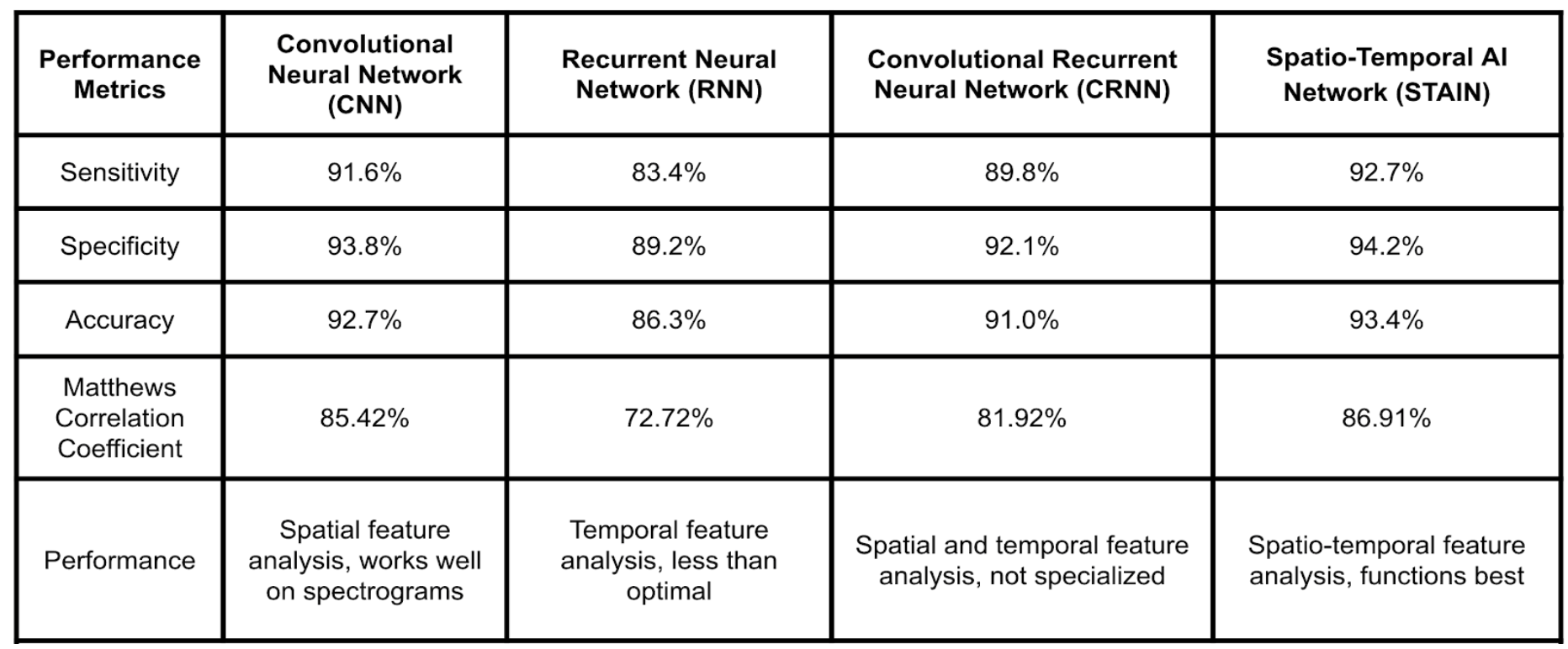}
      \caption{Summary of results of the comparative study of CNN, RNN, CRNN, and the proposed STAIN machine learning models for cough detection. The sensitivity, specificity, accuracy, and Matthews Correlation Coefficient metrics were obtained for all four models using the same datasets. As can be seen, the STAIN model outforms all the other traditional AI models with it’s deeply meshed spatio-temporal feature extraction architecture, which is more advantageous for effectively classifying respiratory events.}
\end{figure*}

\begin{figure*}[thpb]
      \centering
      \label{fig3}
      \includegraphics[width = 6in]{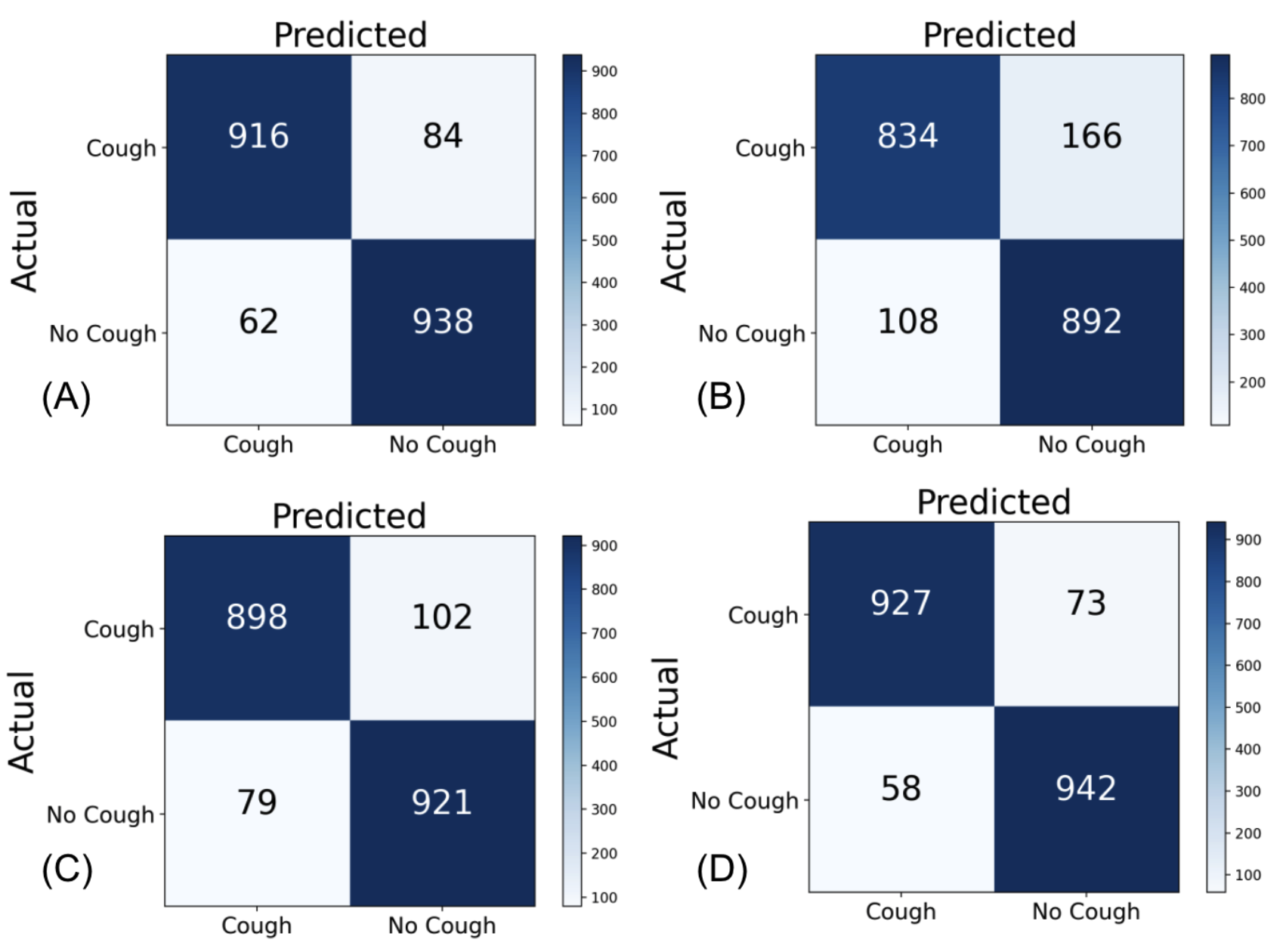}
      \caption{Confusions Matrices, created using MatPlotLib, for (A) CNN, (B) RNN, (C) CRNN, and (D) the proposed STAIN machine learning models. The new STAIN architecture outperforms the traditional neural network architectures for more accurate cough detection.}
\end{figure*}

Using the dataset described in the previous section, rigorous evaluations of the four different AI models were performed. The results of these analyses are shown in \hyperref[table2]{Fig. 4} and \hyperref[fig3]{Fig. 5}, which present the following performance metrics: sensitivity, specificity, accuracy, Matthews Correlation Coefficients, and the confusion matrices.

As these results illustrate, compared to RNN’s temporal feature analysis, CNN’s spatial analysis was better suited for classifying spectrograms. CRNN, created by simply stacking the CNN and RNN components, could not bring out the best of both architectures, performing worse than CNN. In contrast, the proposed new machine learning model, STAIN, performed better than all other models using its architecture for deeply meshed spatio-temporal feature analysis.

\begin{figure*}[thpb]
      \centering
      \label{fig4}
      \includegraphics[width = 6in]{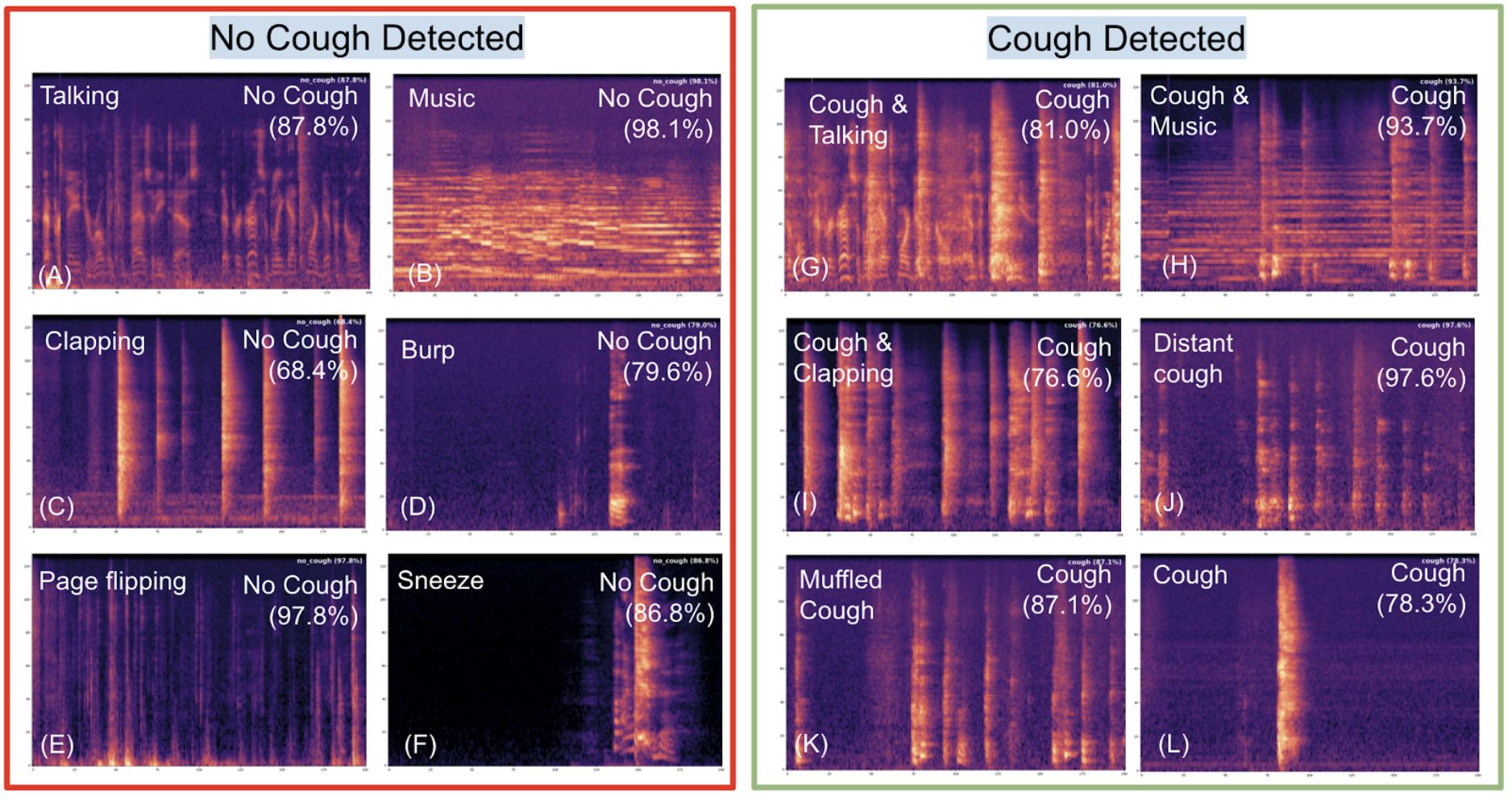}
      \caption{Screenshots of the live demonstration of the cough detection module based on the new spatio-temporal machine learning model. The real-time application, implemented on a laptop computer, captures user-generated sounds using it’s integrated microphones, converts the sound into spectrogram images, processes through the STAIN model to detect the presence of cough, and displays the results on the screen.}
\end{figure*}

\begin{figure*}[thpb]
      \centering
      \label{fig4}
      \includegraphics[width = 6in]{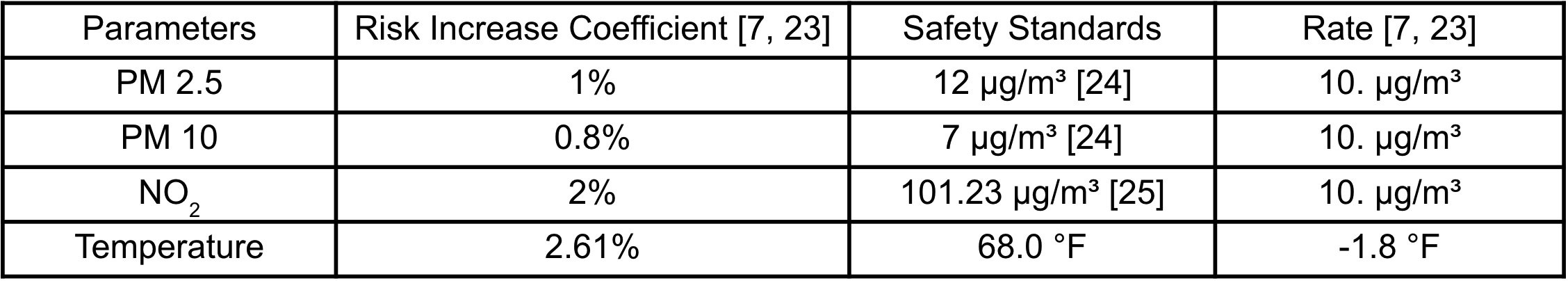}
      \caption{Correlations between the degradation of the environmental and meteorological factors and the increase in COPD exacerbation risks, derived from retrospective medical studies [7, 23]. As an example, these studies demonstrated that an increase in NO$_{2}$ concentration by 10 ug/m³ resulted in about 2\% increase in the risk. These correlations were used to estimate the overall risk trends based on the real-time data from local sensors.}
\end{figure*}

\begin{figure*}[thpb]
      \centering
      \label{eq1}
      \includegraphics[width = 6in]{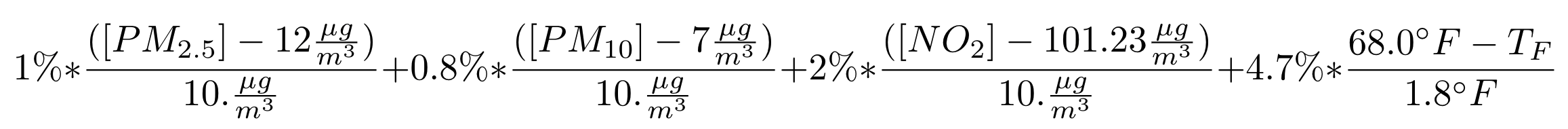}
      \caption{Equation to estimate the increase in COPD exacerbation risks as a function of environmental and meteorological factors (PM$_{2.5}$, PM$_{10}$, NO$_{2}$, and T$_\text{F}$ for Temperature), derived based on the retrospective medical studies [7, 23].}
\end{figure*}

\subsubsection{Demonstration of the Detection Module}

A live demo application for the real-time cough detection module has been developed. This application, running on a laptop computer, captures user-generated sounds using the built-in microphones of the computer, converts the sound files into spectrogram images, processes the data through the STAIN machine learning model, classifies and tracks the cough count and cough frequency over time. The results are presented on the computer screen with a live display of the spectrogram images corresponding to the sound, superimposed with the classification results of the cough events.
\hyperref[fig4]{Fig. 6} shows the representative screenshots of the application running real-time, and correctly classifying talking, clapping, page flipping, music, burp, and sneezes as “No Cough” (left-hand side of \hyperref[fig4]{Fig. 6}), whereas successfully detecting cough events superimposed with the same background sound environments (right-hand side of \hyperref[fig4]{Fig. 6}).
 
\begin{figure*}[thpb]
      \centering
      \label{fig5}
      \includegraphics[width = 6in]{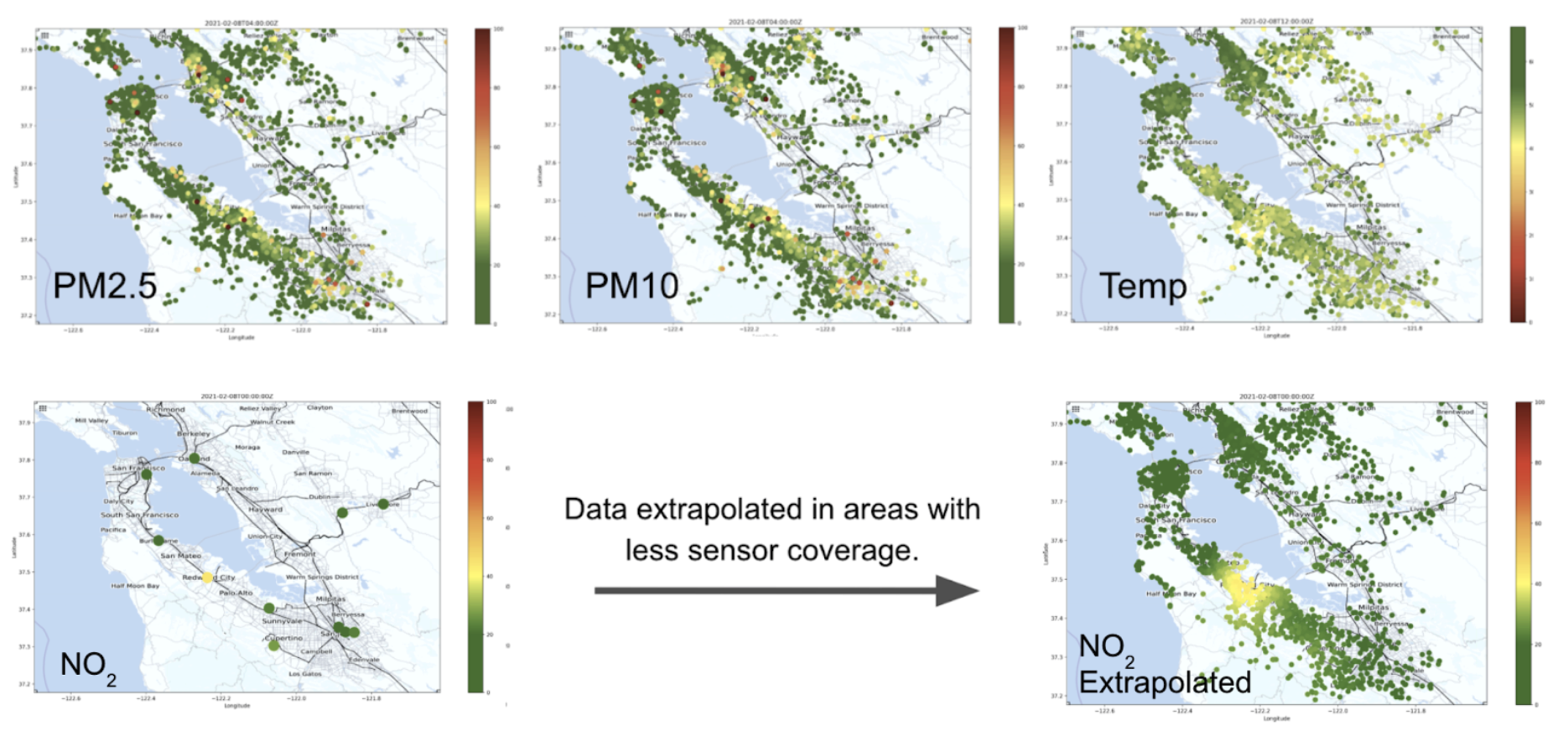}
      \caption{Data maps for the relevant environmental and meteorological factors (PM$_{2.5}$, PM$_{10}$, NO$_{2}$, and Temperature), obtained from the sensors deployed by PurpleAir and the WAQI data platform. An extrapolation method was used to estimate the data in areas with sparse sensor coverage.}
\end{figure*}

\begin{figure*}[thpb]
      \centering
      \label{fig6}
      \includegraphics[width = 6in]{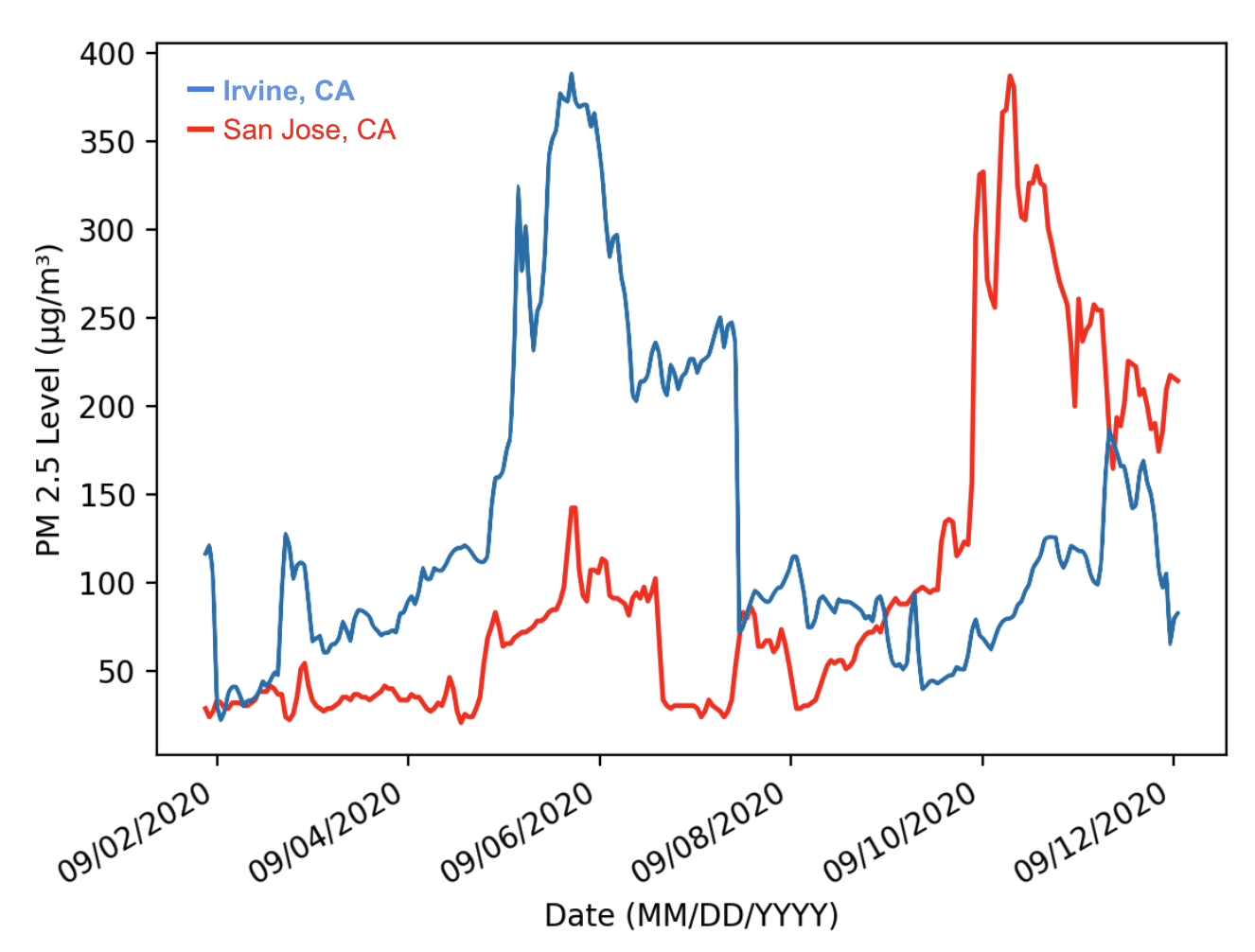}
      \caption{Sensors deployed by PurpleAir in Irvine and San Jose showed that the PM$_{2.5}$ concentration spiked to dangerous levels during Sept. 2-13, 2020, fire season. The onsets of spikes on Sept 6 and Sept 10 correspond to the El Dorado Fire and the SCU Lightning Fire events.}
\end{figure*}

\begin{figure*}[thpb]
      \centering
      \label{fig7}
      \includegraphics[width = 6in]{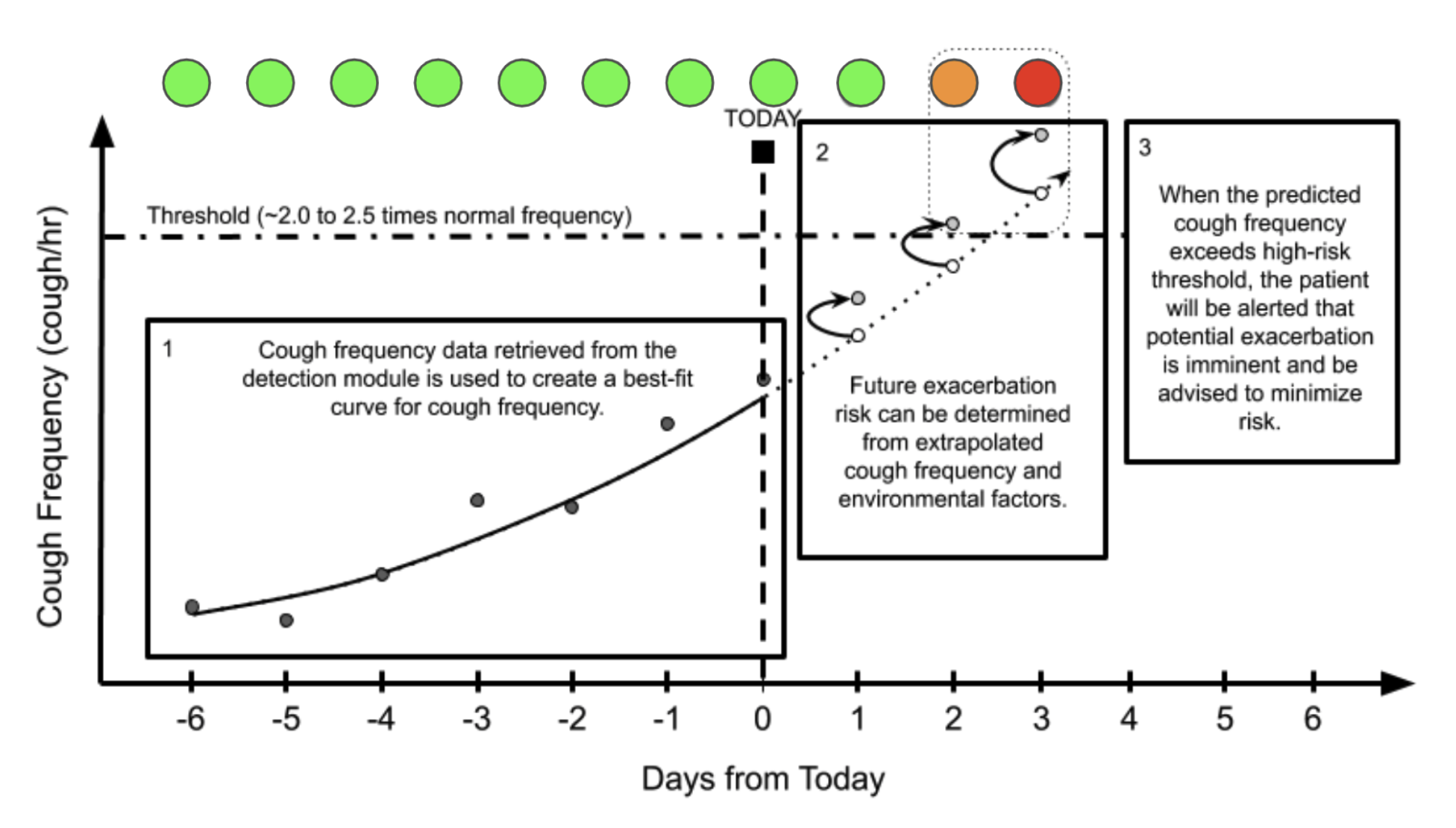}
      \caption{Illustration of the procedures implemented within the prediction module that forecasts the expected progression of the condition of the patient in the days ahead. This final step in the multi-modal architecture combines the results from the respiratory sound analysis performed by the machine learning model of the detection module, and the environmental and meteorological factors and trends analysis conducted by the environmental module. By extrapolating the cough frequency trends, along with the predicted exacerbation risks due to the environmental and meteorological data, the system can alert the patient and caregivers of the imminent risks and preempt medical interventions to potentially reduce hospitalization costs.}
\end{figure*}
 
\subsection{Environmental Module}

While the detection module presented in the previous section tracks real-time cough frequency for patient-specific analysis, the environmental module offers local area-wide environmental and meteorological factor analysis. By examining certain environmental indicators, a patient’s increase of COPD exacerbation likelihood can be determined.

Breathing air quality is one of the most crucial factors in human health; poor air quality can cause any person’s health to significantly deteriorate and is an increasingly important issue following the advent of rapid industrialization. Especially since their lungs are compromised due to inflammation, COPD patients are extremely susceptible to exacerbations caused by bad air quality. A seminal retrospective study analyzed hospitalization and exacerbation rates for COPD patients as functions of the local environmental and meteorological factors, including the concentration of fine particulate matters (where PM$_{x}$ refers to particles or droplets present in the air that are $x$ micrometers or less in width), NO$_{2}$, and temperature variations [7, 23]. These medical studies established that the percentage exacerbation risk increases are directly proportional to PM$_{2.5}$ and PM$_{10}$ levels, NO$_{2}$ concentrations, and temperature variations. The details of the findings are outlined in \hyperref[table3]{Fig. 7}, with each increase/decrease of the “Rate” from “Safety Standards” constituting an additional “Risk Increase Coefficient” for exacerbations.

Based on the results of these retrospective medical studies, an equation has been formulated in this project to estimate the percentage exacerbation risk increase using the four environmental and meteorological parameters in the patient’s location, as shown in \hyperref[eq1]{Fig. 8}. If a factor falls below the threshold standard, its contribution to the final risk percentage is zero; otherwise, it follows the  formula outlined in \hyperref[eq1]{Fig. 8}.

In order to generate a real-time risk map that would represent the exacerbation risk increase for an individual given the environmental factors in the patient’s location, the environmental and climatological data measured by sensors deployed by PurpleAir which are accessible via an open-source database [26], and NO$_{2}$ readings from the World Air Quality Index (WAQI) data platform [27], have been incorporated into the above equation and overlaid on the geographical map of the region. Moreover, an extrapolation method has been developed to estimate the data at a specific location using the data from the sensors deployed in adjacent areas. As an example, \hyperref[table4]{Fig. 9} shows the data map for PM$_{2.5}$, PM$_{10}$, Temperature, and NO$_{2}$ from over 6000 sensors in the San Francisco Bay Area. As a spot check for the data, \hyperref[fig5]{Fig. 10} shows the PM$_{2.5}$ concentrations recorded by the PurpleAir sensors in Irvine and San Jose areas during the first half of September, 2020. The onsets of spikes on Sept 6 and Sept 10 correspond to the El Dorado Fire and the SCU Lightning Fire events, respectively.

\subsection{Prediction Module}
Finally, the prediction module combines the results of the respiratory sound analysis from the detection module and the environmental and meteorological factors analysis from the environmental module to forecast a patient’s expected conditions.

Previously reported medical research studies have determined average cough frequencies for COPD-affected smokers, affected ex-smokers, healthy smokers, and healthy non-smokers [24, 25]. Thus, by extrapolating the progression in cough frequency as determined by the STAIN machine learning model and exacerbation risk increase from environmental factors from the data trends, a patient's expected condition is determined.

This method is illustrated in \hyperref[fig6]{Fig. 11}. First, based on the continuous respiratory event classifications performed by the STAIN machine learning model within the detection module, a best-fit curve is created to determine the patient’s cough frequency trend. Next, the future exacerbation risks are derived based on the extrapolated cough frequency data and the increased risks due to environmental and meteorological factors as determined by the correlations established by the retrospective medical studies, as explained in the previous section. If the prediction module forecasts exceeding the threshold levels that are acceptable, the patient and caregivers would be alerted of the imminent exacerbations for necessary early medical interventions, thereby improving the patient’s quality of life and potentially saving hospitalization costs.

\section{\textbf{CONCLUSION}}
In summary, a multi-modal technique has been developed for predicting the exacerbation risks for respiratory diseases such as COPD, based on a new artificial intelligence model for respiratory sound analysis and retrospective medical studies correlating key environmental parameters to exacerbations. The proposed solution includes a novel spatio-temporal machine learning model for accurate real-time classification and monitoring of respiratory conditions, tracking of local environmental and meteorological factors with commercially deployed sensors, and forecasting the patient’s progression of conditions by combining the trends derived from these two modules.

The proposed new spatio-temporal artificial intelligence network architecture deeply meshes the salient structures of both convolutional and recurrent neural networks, and as a result outperforms both traditional CNN and RNN models, as well as the more recent CRNN models, in extracting the spatial and temporal features that are inherent in spectrograms of respiratory sounds. Extensive comparative tests have been performed to demonstrate that the new model achieves better sensitivity, specificity, accuracy, and Matthews Correlation Coefficient metrics than the traditional machine learning models.

A telehealth solution based on this work can assess the exacerbation risks and alert patients and doctors of early medical intervention, medication, and impending hospitalization. Thus, this technique can conveniently and cost-effectively help minimize and mitigate the impact of respiratory exacerbations, therefore  improving patients’ quality of life and potentially reducing hospitalization costs.

The future work will include collaboration with medical research institutions to further validate and deploy a remote patient monitoring solution into the real-world.





\section*{\textbf{ACKNOWLEDGMENT}}

I would like to express sincere gratitude to my advisor, Chris Spenner, for providing insightful advices. I would also like to thank Kailas Vodrahalli, Archelle Georgiou, Sridhar Nemala, and Krishna Vastare for their inputs and guidance. Additionally, I would like to thank my parents for their continuous support.


\end{document}